**Disparities in Dermatology AI Performance on a Diverse, Curated Clinical Image Set**

Roxana Daneshjou[1,2]*, Kailas Vodrahalli[3]*, Roberto A Novoa[1,4], Melissa Jenkins[1], Weixin Liang[5], Veronica Rotemberg[6], Justin Ko[1], Susan M Swetter[1], Elizabeth E Bailey[1], Olivier Gevaert[2], Pritam Mukherjee[2], Michelle Phung[1], Kiana Yekrang[1], Bradley Fong[1], Rachna Sahasrabudhe[1], Johan A. C. Allerup[1], Utako Okata-Karigane[7], James Zou[2,3,5,8+], Albert Chiou[1+]

*These authors contributed equally
+These authors contributed equally

Correspondence: jamesz@stanford.edu, achiou@stanford.edu

1. Department of Dermatology, Stanford School of Medicine, Redwood City, CA
2. Department of Biomedical Data Science, Stanford School of Medicine, Stanford, CA
3. Department of Electrical Engineering, Stanford University, Stanford, CA
4. Department of Pathology, Stanford School of Medicine, Stanford, CA
5. Department of Computer Science, Stanford University, Stanford, CA
6. Dermatology Service, Memorial Sloan Kettering Cancer Center, New York, NY, USA
7. Department of Dermatology, Keio University School of Medicine, Tokyo, Japan
8. Chan-Zuckerberg Biohub, San Francisco, CA, USA.

**One sentence summary:** Previously developed artificial intelligence algorithms for identifying skin cancer show bias in performance across diverse skin images; a new diverse skin images benchmark can help mitigate these biases through finetuning.

## Abstract

Access to dermatological care is a major issue, with an estimated 3 billion people lacking access to care globally. Artificial intelligence (AI) may aid in triaging skin diseases. However, most AI models have not been rigorously assessed on images of diverse skin tones or uncommon diseases. To ascertain potential biases in algorithm performance in this context, we curated the Diverse Dermatology Images (DDI) dataset—the first publicly available, expertly curated, and pathologically confirmed image dataset with diverse skin tones. Using this dataset of 656 images, we show that state-of-the-art dermatology AI models perform substantially worse on DDI, with receiver operator curve area under the curve (ROC-AUC) dropping by 27-36% percent compared to the models' original test results. All the models performed worse on dark skin tones and uncommon diseases, which are represented in the DDI dataset. Additionally, we find that dermatologists, who typically provide visual labels for AI training and test datasets, also perform worse on images of dark skin tones and uncommon diseases compared to ground truth biopsy annotations. Finally, fine-tuning AI models on the well-characterized and diverse DDI images closed the performance gap between light and dark skin tones. Moreover, algorithms fine-tuned on diverse skin tones outperformed dermatologists on identifying malignancy on images of dark skin tones. Our findings identify important weaknesses and biases in dermatology AI that need to be addressed to ensure reliable application to diverse patients and diseases.

## Introduction

Globally an estimated 3 billion people have inadequate access to medical care for skin disease.[1] Even in developed countries, such as the United States, there is a shortage and unequal distribution of dermatologists, which can lead to long wait times for skin evaluation.[2] Artificial intelligence (AI) diagnostic and decision support tools in dermatology, which have seen rapid development over the last few years, could help triage lesions or aid non-specialist physicians in diagnosing skin diseases.[3-5] There are several commercial skin cancer detection algorithms with the CE mark in Europe.[6]

Despite the widespread interest in dermatology AI, systematic evaluation of state-of-the-art dermatology AI models on independent real-world data has been limited. Most images used to train and test malignancy identification algorithms use siloed, private clinical image data; the data sources and curating methods are often not clearly described.[5] Moreover, limitations in the current training and testing of dermatology AI models may mask potential vulnerabilities. Many algorithms are trained or tested on the International Skin Imaging Collaboration (ISIC) dataset, which contains histopathology-confirmed dermoscopic images of cutaneous malignancies but lacks images of inflammatory and uncommon diseases, or images across diverse skin tones.[7-9] Online atlases such as Fitzpatrick 17k have unreliable and unknown label sources that do not have histopathological confirmation of malignancies.[10] In fact, no public skin disease AI benchmarks have images of biopsy-proven malignancy on dark skin.[11]

Label noise is also a major concern, as many previously published AI algorithms rely on images labeled by visual consensus – meaning that dermatologists provide labels by reviewing only a digital image without information on follow-up or biopsy confirmation.[11] Visual inspection, however, can be unreliable for determining cutaneous malignancies, which often require histopathological confirmation.[12]

**Results**

**Diverse Dermatology Images Dataset**
To ascertain potential biases in algorithm performance in this context, we curated the Diverse Dermatology Images (DDI) dataset—a pathologically confirmed benchmark dataset with diverse skin tones. The DDI was retrospectively selected from reviewing histopathologically-proven lesions diagnosed in Stanford Clinics from 2010-2020. For all lesions, the Fitzpatrick skin type (FST), a clinical classification scheme for skin tone, was determined using chart review of the in-person visit and consensus review by two board-certified dermatologists. This dataset was designed to allow direct comparison between patients classified as FST V-VI (dark skin tones) and patients with FST I-II (light skin tones) by matching patient characteristics. There were a total of 208 images of FST I-II (159 benign, 49 malignant), 241 images of FST III-IV (167 benign, 74 malignant), and 207 images of FST V-VI (159 benign and 48 malignant) (Supplemental Table 1).

**Previously developed Dermatology AI algorithms perform worse on dark skin tones and uncommon diseases**
We evaluated three algorithms on their ability to distinguish benign versus malignant lesions: ModelDerm (using the API available at https://jid2020.modelderm.com/)[13] and two algorithms with state-of-the-art performance developed from previously described datasets – DeepDerm[4] and HAM 10000[7]. These algorithms were selected based on their popularity, availability, and previous demonstrations of state-of-the-art performance.[4,7,13]

Though all three algorithms had good performance on the original datasets they were trained and tested on, their performance was worse on DDI. In the literature, ModelDerm had a previously reported receiver operator curve area under the curve (ROC-AUC) of 0.93-0.94[13], while DeepDerm achieved ROC-AUC of 0.88 and HAM 10000 achieved ROC-AUC of 0.92 on its own test data. However, when evaluated on the DDI dataset, ModelDerm had a ROC-AUC of 0.65 (95% CI 0.61-0.70), DeepDerm had a ROC-AUC of 0.56 (0.51-0.61), and HAM 10000 had a ROC-AUC of 0.67 (95% CI 0.62-0.71) (Figure 1A, Supplemental Table 2).

We assessed differential performance between FST I-II and FST V-VI; these two subsets of images were matched for diagnostic class (malignant vs. benign) and patient demographics (Supplemental Methods), enabling a direct comparison of performance. Across all three algorithms on the DDI dataset, ROC-AUC performance was better on the subset of Fitzpatrick I-II images (Figure 1B) compared to Fitzpatrick V-VI (Figure 1C) with ModelDerm having a ROC-AUC of 0.64 (FST I-II) versus 0.55 (FST V-VI), DeepDerm having a ROC-AUC of 0.61 (FST I-II) versus 0.50 (FST V-VI), and HAM 10000 having a ROC-AUC of 0.72 (FST I-II) versus 0.57 (FST V-VI). The use of robust training methods on DeepDerm did not reduce the gap in performance between FST I-II and FST V-VI (Supplemental Table 3).[14-16]

Because detecting malignancy is an important feature of these algorithms and clinical care, we assessed sensitivity to see if there was differential performance across skin tones (Supplemental Table 2). Across the entire DDI dataset, two algorithms showed statistically significant differential performance in detecting malignancies between FST I-II and FST V-VI: ModelDerm (sensitivity 0.41 vs 0.12, Fisher's Exact Test p = 0.0025) and DeepDerm (sensitivity 0.69 vs 0.23, Fisher's Exact Test p = $5.65 \times 10^{-6}$). HAM 10000 had poor sensitivity across all subsets of the data despite having a sensitivity of 0.68 on its own test set (Supplemental Table 2). The thresholds determined for each model were based on the maximum F1 score of the original test set used to train each respective algorithm, suggesting that the threshold set for HAM10000 generalized poorly.

The datasets used to train all three AI models consist mostly of common malignancies, while the DDI dataset also includes uncommon benign and malignant lesions. To assess how this distribution shift could contribute to the model's performance drop-off, we repeated our analysis on only the common disease split of the DDI dataset. While removing uncommon diseases led to overall improvement in performance in distinguishing between benign versus malignant disease across all algorithms with ModelDerm having a ROC-AUC of 0.74 (CI 0.69-0.79), DeepDerm having a ROC-AUC of 0.64 (95% CI 0.58-0.69) and HAM 10000 having a ROC-AUC of 0.71 (0.66-0.76), this performance was still lower than performance on each AI algorithm's original test set (Figure 1D). DeepDerm and HAM10000 continued to have a difference in performance between FST I-II and FST V-VI even among common diseases (Supplemental Table 2).

**Fine-tuning on diverse data can close performance gap between light and dark skin tones**

We assessed how using DDI for model fine-tuning could be used to help close the differential performance between FST I-II and FST V-VI for DeepDerm and HAM 10000, the two algorithms for which we had access to model weights. Fine tuning improved overall model performance across all the skin tones. Importantly, we found that the inclusion of the diverse DDI data with fine tuning closed the gap in performance between FST I-II and FST V-VI (Figure 2) for both DeepDerm and HAM 10000. After fine tuning the DeepDerm algorithm on DDI data, we achieved a ROC-AUC of 0.72 (95% CI 0.69-0.76) on FST I-II compared to 0.74 (0.69-0.80) for FST V-VI.  Similarly, after fine tuning the HAM 10000 algorithm, we achieved ROC-AUCs of 0.74 (0.72-0.80) on FST I-II data and 0.77 (0.74-0.80) on FST V-VI data.  Fine-tuning also made the performance of the algorithms to be equivalent or exceeding that of the dermatologists (Figure 2).

**Dermatologist consensus labeling can lead to differential label noise across skin tones**
Many dermatology AI algorithms rely on data labels generated from dermatologists reviewing images alone, a task that is different from actual clinical practice where in-person assessment and diagnostic tests aid diagnosis. To assess the potential for label noise, we compare dermatologist generated labels (benign or malignant) versus biopsy-proven labels across DDI. Notably, dermatologist labels generally outperformed algorithms prior to fine-tuning (Figure 1). Dermatologist labels were less noisy on images of DDI common diseases compared to the whole DDI dataset (ROC-AUC 0.82 to 0.72, sensitivity 0.88 to 0.71, p=0.0089). Label noise also varied between FST I-II and FST V-VI (sensitivity 0.72 vs. 0.59, Fisher's Exact Test p = $8.8 \times 10^{-6}$) and this difference remained statistically significant when assessing only common disease (sensitivity 0.93 versus 0.62, Fisher's Exact Test p = 0.0096).

**Discussion**

Dermatology AI algorithms have been envisioned as providing support to non-dermatologist specialists or helping with triaging lesions prior to clinical care.[17] Our analyses highlight several key challenges for AI algorithms developed for detecting cutaneous malignancies: 1) state-of-the-art dermatology AI algorithms AI algorithms have substantially worse performance on lesions appearing on dark skin compared to light skin using biopsy-proven malignancies, the gold standard for disease annotation; 2) as a consequence, there is a substantial drop-off in the overall performance of AI algorithms developed from previously described data when benchmarked on DDI; however fine-tuning can help close performance gaps between skin tones and 3) there are differences in dermatologist visual consensus label performance, which is commonly used to train AI models, across skin tones and uncommon conditions.

In the current dermatology workflow, there are significant disparities in skin cancer diagnosis and outcomes, with patients with skin of color getting diagnosed at later stages, leading to increased morbidity, mortality and costs.[18,19]  In order to alleviate disparities rather than exacerbate them, dermatology datasets used for AI training and testing must include dark skin tones. We release the DDI benchmark as a first step towards this goal and demonstrate how such data can be useful in further evaluation of previously developed algorithms. Because every lesion in DDI is biopsied, we are less likely to have label noise. The dataset also includes ambiguous benign lesions that would have been difficult to label visually but are representative of the type of lesions seen in clinical practice.

Resources such as DDI can show the limitations of current dermatology AI models, which can be sensitive to its training data distributions.  ModelDerm was trained on predominantly clinical images, DeepDerm used a mix of clinical and dermoscopic images while HAM 10000 only trained dermoscopic images.  ModelDerm has greater skin tone diversity, since it was trained on a mix of white and Asian patients, while DeepDerm and HAM 10000 were trained on predominantly white patients.  Unlike DeepDerm and HAM 10000, ModelDerm did not show a drop-off in ROC-AUC between Fitz I-II and Fitz V-VI on the DDI common diseases dataset.

Interestingly, the HAM 10000 data is more limited in the number of diagnoses compared to DeepDerm or ModelDerm; however, unlike those datasets, every malignancy in HAM 10000 is histopathologically confirmed, meaning there is less label noise.[4,7,13] This could be one reason why the model trained on HAM 10000 achieved the best ROC-AUC performance on the DDI dataset. Moreover, the use of robust training methods that are designed to improve algorithmic fairness to train DeepDerm did not reduce disparity across the skin tones, further suggesting that the performance limitations lie with the lack of diverse training data. The sensitivity-specificity threshold for all the algorithms was set based on the original training/test data to avoid overfitting and to mimic how thresholds are set for real world commercial applications. These thresholds may not generalize well to new datasets; for example, despite a good ROC-AUC, HAM 10000 had very poor sensitivity on the DDI dataset and classified most lesions as benign.

Many previously published algorithms have relied on visual consensus labels from a small group of dermatologists.[4,5,20] However, dermatologist labels from visual inspection can be noisy since they lack pertinent information used for diagnosis such as clinical history, in-person evaluation, and other diagnostic tests. Consistent with our findings, Hekler et al showed that a melanoma classifier trained on dermatologist visual consensus labels rather than ground truth pathologic diagnosis performed significantly worse than a classifier trained on ground truth labels.[21] We found that consensus label performance was worse when uncommon diagnoses were included and on FST V-VI skin tones. Biases in photography such as color balancing could be a contributor as could the difficulty of capturing salient features such as erythema in photography of dark skin tones.[22] Prior studies have also suggested that the lack of diverse skin tones in dermatology educational materials may also contribute.[23-25] Here, we make no claims regarding overall general dermatologists' performance, but illustrate the potential for differences in label noise across skin-tones during the consensus labeling process.

Prior to the DDI dataset, there were no publicly available benchmark datasets that included biopsy-proven malignancies in dark skin.[11] Though this dataset increases the number of biopsy-proven diagnoses across diverse skin tones, it has limitations due to the nature of biopsy-curated data. Because benign lesions are not regularly biopsied, this dataset is enriched for "ambiguous" lesions that a clinician determined required biopsy; however these lesions are reflective of the kinds of lesions an algorithm may encounter. This dataset is not comprehensive of all diagnoses in dermatology and is tailored towards algorithms for triaging malignant from benign lesions, a common task in dermatology AI.[4,11,26] primary analyses focus on comparing FST I-II and FST V-VI images because they were matched across diagnostic category, patient age, sex and date of photos and because they exhibit the most significant performance disparities. While this is not comprehensive of all the potential confounders that could exist when comparing FST I-II and V-VI, our fine-tuning experiments show the gap in performance can be closed by using diverse data. Although FST is used most commonly for labeling skin tones for images used in AI studies, this scale has limitations and does not capture the full diversity of human skin tones.[27]

While the DDI dataset may not be sufficiently large for training models from scratch, we found that differences in performance between FST I-II and FST V-VI in previously state-of-the-art algorithms could be overcome by fine-tuning on DDI. Interestingly, the fine-tuned algorithms performed better than the dermatologist labelers on skin tones V-VI, suggesting that future algorithms trained on diverse data could have the potential for providing decision support. Thus, we believe this dataset will not only provide a useful evaluation benchmark but also allow

model developers to reduce the performance disparity across skin-tones in dermatology AI algorithms

**Materials and Methods**

**Image selection and processing**
The Stanford Research Repository (STARR) was used to identify self-described Black, Hispanic, or Latino patients who received a biopsy or tangential shave (CPT codes: skin biopsy 11100, tangential shave 11102) within the past 10 years starting from Jan 1, 2010. Per institutional practice, all biopsied lesions undergo clinical photography taken from approximately 6 inches from the lesion of interest, typically using a clinic-issued smartphone camera. These clinical images were pulled from the electronic medical record. Lesions were characterized by specific histopathologic diagnosis, benign vs. malignant status, date of biopsy, and age, sex, and Fitzpatrick skin type (FST) of the patient. A patient's FST was labeled using clinical documentation from an in-person visit, demographic photo if available, and the image of the lesion. Two board-certified dermatologists (RD and AC) performed final review using a consensus process to adjudicate any discrepancies. Discrepancies between the label in the clinical note and the labelers were seen in less than 1% of the data; there was full agreement between the consensus labelers in those cases.

FST I/II patients were generated by matching each lesion diagnosed in the FST V/VI cohort with a lesion with the identical histopathologic diagnosis diagnosed in a patient with FST I/II regardless of race using Powerpath software. Patient demographics were assessed using chart review and included as a matched control if the lesion was diagnosed within the same 3 year period to account for incremental improvements in phone camera technology, and if the patient had the corresponding sex, age within 10 years, and FST type I/II. FST was confirmed in the same manner as the prior cohort. In instances where a matching control strictly meeting the above criteria could not be identified, a lesion with a similar diagnostic category (i.e. matching one type of non-melanoma skin cancer with another type) in an FST I/II patient meeting the majority of the control criteria would be included instead to preserve the ratio of malignant to benign lesions. During this process, skin lesions associated with patients determined to be FST III/IV were included in the overall FST III/IV group, which represents a convenience sample that was not matched to the FST V/VI group.

Diagnosis labels and malignant versus benign labels were determined by a board-certified dermatologist (RD) and board-certified dermatopathologist (RN) reviewing the histopathologic diagnoses. Any additional tests ordered on pathology was also considered for borderline cases – for example, atypical compound melanocytic proliferation with features of an atypical spindle cell nevus of Reed was labeled as malignant due to diffuse loss of p16 on immunohistochemical stains, while an atypical lymphocytic infiltrate was confirmed to be benign based on negative clonality studies.

**Released images**
Prior to image release, an additional review of all photos was done. Additional cropping was done after independent reviews by four board-certified dermatologists as part of the expert determination process of ensuring that all images were de-identified (no full face images, identifiable tattoos, unique jewelry, or labels with potential identifiers). This was a Stanford IRB approved study: protocol numbers 36050 and 61146.

**Dermatologist reader study and quality filtering**
Three independent board-certified dermatologists with 5, 9, and 27 years of experience and no previous access to the image data or pathology labels were asked to assess each image in an untimed manner. They were asked to assign a photo quality score using a previously developed image quality scale.[28] Overall quality scores were tabulated by taking the mean of all three dermatologists. Thirteen images did not meet quality metrics and were removed from analysis, leaving a total of 656 images. We compared the quality score distribution between Fitzpatrick I-II photo and Fitzpatrick V-VI photos using the Mann Whitney U test, showing no significant difference in quality (scipy.stats package, ipython 7.8.0). Additionally, dermatologists were asked to rate whether they thought the presented image showed a benign or malignant process. An ensemble of the three dermatologists' predictions was created using a majority vote.

**AI algorithms**
The labels generated by AI algorithms came from three sources, labeled "ModelDerm" and "DeepDerm", and "HAM 10000".

ModelDerm is a previously described algorithm with an available online API (https://jid2020.modelderm.com/); outputs were generated in December 2020.[13]

DeepDerm is an algorithm developed from using previously described data and similar parameters to the algorithm developed by Esteva *et al*.[4] The algorithm uses the Inception V3 architecture.[24] The data sources include ISIC, two datasets sourced from Stanford hospital, Edinburgh Dermofit Library, and open-access data available on the web.[4] We mix all datasets and randomly generate train (80%) and test splits (20%). Images are resized and cropped to a size of 299 x 299 pixels. During training, we also augment data by randomly rotating and vertically flipping images before the resize and crop operation previously described; the rotation operation involves rotating the image and cropping to the largest upright, inscribed rectangle in the image. We train using the Adam optimization method with a learning rate of $10^{-4}$ and with binary cross entropy loss to classify images as malignant / benign.[29] We use balanced sampling across benign and malignant images during training.

HAM 10000 is an algorithm developed from the previously described, publicly available HAM 10000 dataset.[7] The algorithm and training methods are identical to the DeepDerm methods, with the only difference being the data used. The HAM 10000 dataset consists of 10,015 dermoscopy images; all malignancies are biopsy confirmed.[7]

We also assessed three robust training methods using DeepDerm's training data: GroupDRO, CORAL, and CDANN, which have been shown to reduce the disparity of AI models.[14-16] These methods require us to partition our dataset into the groups we would like to be robust across. We use the source of each image to define the groups, since the data source can capture confounders, artifacts as well as imbalances across skin color that we would like the AI model to be robust to. As we have 5 source datasets and 2 classes (benign and malignant), we define 10 groups in total (5 sources x 2 classes).

Each of these robust training methods is then trained using the same data augmentation strategies and optimization algorithm as described above, with the exception that we perform balanced sampling across these 10 groups. GroupDRO minimizes the maximum expected loss across the 10 groups to be robust to the worst-case training loss across groups; it additionally adds a regularization penalty to be more effective in the deep learning setting. CORAL minimizes the standard cross entropy loss but adds an additional penalty to force the intermediate image embedding to be similar across groups. In particular, CORAL penalizes differences in mean and covariance of the image embedding between all group pairs. CDANN also attempts to enforce domain-invariant image embeddings but uses a discriminator network instead that is trained to distinguish between domains. The loss from this network forces embeddings to be similar across domains. We train these robust training algorithms across 5 seeds and calculate the mean and 95% confidence interval ROC-AUC values.

**Data analysis**
We performed each analysis with all 656 images, which included a mix of common and uncommon skin diseases. To assess the effects of uncommon disease on performance when assessing benign versus malignant lesions, we removed any diagnosis that was reported in the literature to have an incidence less than 1 in 10,000 in the entire population or in the absence of relevant literature, were determined to be uncommon by three board certified dermatologists and dermatopathologists.[30-38] Consistent with a previously reported taxonomy, we considered diseases as a whole and not based on subtypes (for example, acral lentiginous melanoma is considered as part of melanoma in considering incidence).[4] Three board certified dermatologists independent from the labelers (RD, RN, AC) assessed the diseases to ensure that diseases labeled as "uncommon" were considered uncommon among dermatologists.

Benign diagnoses that were considered common included: Abrasions, ulcerations, and physical injuries; Abscess; Acne (cystic); Acral melanotic macule; Acrochordon; Actinic keratosis; Angioma; Atypical lymphocytic infiltrate; Benign keratosis; Blue nevus; Cherry angioma; Clear cell acanthoma; Condyloma acuminatum; Dermatofibroma; Dysplastic nevus; Eczema/spongiotic dermatitis; Epidermal cyst; Epidermal nevus; Fibrous papule; Folliculitis; Foreign body granuloma; Hematoma; Hyperpigmentation; Keloid; Lichenoid keratosis; Lipoma; Melanocytic nevi; Molluscum contagiosum; Neurofibroma; Neuroma; Onychomycosis; Prurigo nodularis; Pyogenic granuloma; Reactive lymphoid hyperplasia; Scar; Seborrheic keratosis; Solar lentigo; Tinea pedis; Trichilemmoma; Verruca vulgaris/wart

Benign diagnoses that were considered uncommon included: Acquired Digital Fibrokeratoma; Angioleiomyoma; Arteriovenous hemangioma; Cellular neurothekeoma; Chondroid syringoma; Cutaneous coccidioidomycosis; Dermatomyositis; Eccrine poroma; Focal acral hyperkeratosis; Glomangioma; Graft-vs-host disease; Inverted follicular keratosis; Morphea; Nevus lipomatosus superficialis; Pigmented spindle cell nevus of Reed; Syringocystadenoma papilliferum; Trichofolliculoma; Verruciform xanthoma; Xanthogranuloma

Malignant diagnoses that were considered common included: Basal cell carcinoma; Melanoma; Melanoma in situ; Squamous cell carcinoma; Squamous cell carcinoma; keratoacanthoma type; Squamous cell carcinoma in situ

Malignant diagnoses that were considered uncommon included: Atypical compound melanocytic proliferation (features of atypical spindle cell nevus of Reed with diffuse P16 loss requiring re-excision); Blastic plasmacytoid dendritic cell neoplasm; Kaposi sarcoma; Leukemia cutis; Metastatic carcinoma (cutaneous metastases); Mycosis Fungoides; Sebaceous carcinoma; Subcutaneous T-cell lymphoma

For calculating Receiver Operator Curve Area Under the Curve (ROC-AUC), we used the probabilities generated by each algorithm. For the Ensemble Dermatologist ROC-AUC, we used probabilities generated by summing the votes and dividing by the total number of dermatologists. We calculated a 95% confidence interval using bootstrapping with 50,000 iterations. We assess for drop-offs in sensitivity between FST I-II and FST V-VI using Fisher's Exact Test to compare the proportion of true positives and false negatives.

Datasets of dermatology images can have extraneous artifacts such as rulers and markings. However, predicting the correct label (benign or malignant) did not correlate with the presence of markings for dermatologists (Pearson's correlation, r= -0.03, p = 0.50), ModelDerm (Pearson's correlation, r = -0.01, p = 0.91), or HAM 10000 (Pearson's correlation, r = -0.01, p = 0.80). DeepDerm performance was negatively correlated with the presence of markings (Pearson's correlation, r = -0.20, p = 3.2 x $10^{-5}$). Because FST V-VI had fewer images with markers and rulers, these artifacts do not explain why DeepDerm performed worse on the FST V-VI dataset.

**Fine-tuning DeepDerm and HAM 10000 using DDI**
We fine-tuned the DeepDerm and HAM10000 algorithms on the DDI data. We were not able to fine-tune ModelDerm since we can only access it through an API. We perform a random 60-20-20 split into train, validation, and test sets. The split is randomized across skin tone and disease classification groups (e.g., so the ratio of malignant Fitz I-II images is the same across all three groups). During training, we augment our dataset using random rotations, random cropping, random color jitter, and random blurring. We also use Mixup data augmentation during training with parameter alpha=1.[39] Mixup generates synthetic data by taking interpolations of pairs of training data and has been shown to improve model generalizability.[39,40] We fine-tune all layers of our model using the Adam optimizer with a learning rate of 5 x $10^{-2}$ and weight regularization of 1 x $10^{-4}$. We train for up to 500 epochs, and use the validation loss to select the model weights we keep. During testing, we select a center crop of the image. This procedure is repeated for 20 random seeds, where we randomly sample the 60-20-20 data split differently for each seed. We report the average test AUC across these 20 seeds along with 95% confidence intervals in Figure 2.

**Data and code availability**
The DDI dataset is available at https://ddi-dataset.github.io. Our trained AI models are available at https://drive.google.com/drive/folders/1WscikgfyQWg1OTPem_JZ-8EjbCQ_FHxm.

**Acknowledgements:**
J.Z. is supported by NSF CAREER 1942926. A.C. is supported by a Dermatology Foundation Medical Dermatology Career Development Award. A.C, R.A.N., J.K, S.M.S., and O.G. are supported by the Melanoma Research Alliance's L'Oreal Dermatological Beauty Brands-MRA Team Science Award. R.D. is supported by 5T32AR007422-38. K.V. is supported by an NSF graduate research fellowship and a Stanford Graduate Fellowship award.

**Author contributions:**
Concept and Design: R.D., K.V., R.A.N., M.J., W.L., V.R., J.M.K., J.Z., A.S.C
Acquisition, analysis, or interpretation of data: All authors
Drafting of manuscript: R.D., K.V., R.A.N., J.Z., A.S.C.
Critical revision of the manuscript for important intellectual content: R.D., K.V., R.A.N., W.L., V.R., J.M.K., S.M.S., E.E.B., J.Z., A.S.C.
Statistical analysis: R.D., K.V., W.L., J.Z.
Obtained funding: R.A.N., J.M.K., S.M.S., J.Z., A.S.C.

Administrative, technical, or material support: J.Z., A.S.C.
Supervision: J.Z., A.S.C.

**Competing interests:** The authors declare no competing interests.

**Supplemental Tables and Figures**

**Supplemental Table 1:** Number of images after quality filtering; number in parenthesis indicates the subset of each category that represent "common" diagnosis in dermatology

|  | Benign | Malignant | Total |
| --- | --- | --- | --- |
| Fitzpatrick I-II | 159 (148) | 49 (42) | 208 (190) |
| Fitzpatrick III-IV | 167 (153) | 74 (65) | 241 (218) |
| Fitzpatrick V-VI | 159 (140) | 48 (16) | 207 (156) |
| Total Images | 485 (441) | 171 (123) | 656 (564) |

**Supplemental Table 2:** Performance of ModelDerm, DeepDerm, HAM 10000 and an ensemble of dermatologists on the entire DDI dataset (656 images) and the subset of common diagnoses only (564 images) as well as stratification by light (FST I-II) vs dark (FST V-VI) skin tones. ROC-AUC for algorithms was calculated using the probability outputs of each respective algorithm.

ROC-AUC for dermatologists was calculated using probabilities generated from ensembling each dermatologist's vote. Sensitivity and specificity of algorithms were calculated by using the cutoffs previously determined for ModelDerm, DeepDerm, and HAM 10000 during algorithmic development on their respective test sets. Sensitivity and specificity values of the ensemble of dermatologists were calculated by using the majority vote as the label.

| Labeler | Dataset | ROC-AUC | | | Sensitivity (Recall) | | | Specificity | | |
| --- | --- | --- | --- | --- | --- | --- | --- | --- | --- | --- |
| | | All | FST I-II | FST V-VI | All | FST I-II | FST V-VI | All | FST I-II | FST V-VI |
| **ModelDerm** | DDI | 0.65 (0.61-0.70) | 0.64 (0.55-0.73) | 0.55 (0.46-0.64) | 0.36 | 0.41 | 0.12 | 0.83 | 0.75 | 0.89 |
| | DDI Common diseases | 0.74 (0.69-0.79) | 0.68 (0.58-0.77) | 0.70 (0.57-0.82) | 0.47 | 0.45 | 0.25 | 0.83 | 0.77 | 0.88 |
| **DeepDerm** | DDI | 0.56 (0.51-0.61) | 0.61 (0.50-0.71) | 0.50 (0.41-0.58) | 0.53 | 0.69 | 0.23 | 0.53 | 0.38 | 0.68 |
| | DDI Common diseases | 0.64 (0.58-0.69) | 0.64 (0.54-0.75) | 0.55 (0.42-0.67) | 0.64 | 0.71 | 0.31 | 0.52 | 0.39 | 0.68 |
| **HAM 10000** | DDI | 0.67 (0.62-0.71) | 0.72 (0.63-0.79) | 0.57 (0.48-0.67) | 0.06 | 0.02 | 0.06 | 0.99 | 0.99 | 0.99 |
| | DDI Common diseases | 0.71 (0.66-0.76) | 0.75 (0.66-0.82) | 0.62 (0.47-0.77) | 0.07 | 0.02 | 0.06 | 0.99 | 0.99 | 0.99 |
| **Dermatologist Ensemble** | DDI | 0.72 (0.68 - 0.77) | 0.75 (0.68-0.81) | 0.62 (0.54-0.71) | 0.71 | 0.84 | 0.40 | 0.67 | 0.60 | 0.79 |
| | DDI Common diseases | 0.82 (0.79-0.86) | 0.80 (0.73-0.85) | 0.81 (0.70-0.90) | 0.88 | 0.93 | 0.62 | 0.67 | 0.61 | 0.79 |

**Supplemental Table 3:** DeepDerm overall performance and performance on FST I-II and FST V-VI after using fairness-aware methods to train DeepDerm algorithm.

| Method | Overall ROC-AUC mean (95% CI) | FST I-II ROC-AUC mean (95% CI) | FST V-VI ROC-AUC mean (95% CI) |
| --- | --- | --- | --- |
| CDANN[16] | 0.59 (0.56-0.63) | 0.61 (0.54-0.68) | 0.48 (0.43-0.53) |
| CORAL[15] | 0.58 (0.54-0.61) | 0.61 (0.56-0.67) | 0.45 (0.42-0.47) |
| GROUP-DRO[14] | 0.60 (0.58-0.62) | 0.63 (0.60-0.65) | 0.50 (0.48-0.53) |